\documentclass[prd,preprint,nofootinbib]{revtex4}

\usepackage{graphicx}
\usepackage{amssymb}
\usepackage{amsmath}
\usepackage{color}

\begin{document}

\begin{flushright}
UT-11-20\\
\end{flushright}


\title{
Domain Walls and Gravitational Waves after Thermal Inflation
}


\author{Takeo Moroi and Kazunori Nakayama}

\affiliation{%
Department of Physics, University of Tokyo, Bunkyo-ku, Tokyo 113-0033, Japan\\
}

\date{\today}

\vskip 1.0cm

\begin{abstract}
  Thermal inflation is an attractive solution to the cosmological
  moduli problem.  However, domain walls may be formed after thermal
  inflation and some mechanisms are needed to eliminate the domain
  wall before it dominates the Universe.  We point out that
  gravitational waves produced by the dynamics of domain walls may be
  observed by the pulsar timing experiments and future space-borne
  gravitational wave detectors, which provides a probe into the period
  of thermal inflation.  We also show that the QCD instanton effect
  can effectively eliminate the domain walls
  with producing observable amount of gravitational waves.
\end{abstract}

\maketitle

\section{Introduction}

While the supergravity and string theory are well-motivated candidates
of the fundamental theory, there often appear light scalar fields,
called moduli, having Planck-scale suppressed interactions with
matter.  The moduli cause serious cosmological problems, since the
lifetime is typically longer than 1~sec and the energy density stored
in the form of coherent oscillation is too much.  The decay of
  moduli usually injects significant amount of energy after the
big-bang nucleosynthesis (BBN) begins, and it modifies subsequent
cosmological scenarios.  This is called the cosmological moduli
problem~\cite{Coughlan:1983ci,Banks:1993en}.  Even if the moduli are
heavy enough to decay before BBN, gravitinos produced by the moduli
decay again may be problematic~\cite{Endo:2006zj}.

Thermal inflation~\cite{Yamamoto:1985rd,Lazarides:1985ja,Lyth:1995hj,Lyth:1995ka}
is a short period of inflation which takes place well after the
primordial inflation.  It is regarded as a solution to the cosmological
moduli problem, since the coherent oscillation of the moduli is
sufficiently diluted by the exponential expansion during thermal
inflation and subsequent entropy-production by the flaton decay.
Thermal inflation is driven by a scalar field, called {\it flaton},
having a flat potential and trapped at the origin of the scalar
potential due to thermal effects.  This kind of scalar field can be
embedded into supersymmetric (SUSY) theories.

However, the most thermal inflation models suffer from a problematic
domain wall (DW) formation after the phase transition in association
with the end of thermal inflation~\cite{Lyth:1995ka}.  This is because
the flaton usually has $Z_n$ symmetry which is spontaneously broken at
the true vacuum.  Once formed, DWs eventually dominate the energy density of
the Universe resulting in a cosmological disaster.  Thus the $Z_n$
symmetry, which guarantees the flatness of the flaton potential, must
be explicitly broken to some extent in order for the DWs to disappear
before they come to dominate the Universe.  Actually, an introduction
of a small explicit breaking in the scalar potential can solve the DW
problem~\cite{Asaka:1997rv,Asaka:1999xd}.  Such DWs can be a source of
gravitational waves (GWs).

In this paper, we study the GWs from the dynamics of DWs produced
  after thermal inflation.  We show that the amount of the GWs may be
  within the reach of the pulsar timing experiments and future
  space-borne GW detectors.  Thus the study of GWs provides a way to
  probe the period of thermal inflation.  It is also shown that the
  $Z_n$ can be naturally broken by the QCD instanton effect in a
  simple class of thermal inflation model; in such a model, the
  abundance of the GWs will be well within the observable range.

\section{Thermal inflation model}

Let us start with the model of thermal
  inflation.  We introduce a gauge singlet superfield $\phi$ which
  takes the role of the flaton.\footnote{ We use a same symbol for a
    superfield as its scalar component.  }  Imposing $Z_n$ symmetry,
  under which $\phi$ transforms as $\phi\rightarrow e^{2\pi i/n}\phi$,
  we adopt the following superpotential\footnote{Terms like
    $\phi^{2n}, \phi^{3n},\dots$ are also allowed, but they have
    little effects on the flaton dynamics discussed below and hence are neglected.   }
\begin{equation}
  W = \frac{\phi^n}{nM^{n-3}} + k \phi Q\bar Q + W_0.
  \label{W}
\end{equation}
Here, $Q$ and $\bar{Q}$ are additional ``quarks'' which have quantum
numbers for SU(3)$_{\rm QCD}$; they are thermalized when $\phi=0$, and
give a thermal mass to $\phi$.  In addition, $W_0$ is a constant which
is related to the gravitino mass $m_{3/2}$ as $W_0 = m_{3/2}M_P^2$
(with $M_P$ being the reduced Planck scale).  In the following, in
order to guarantee the flatness of the potential of $\phi$, we take $n
\geq 4$.  We also comment here that, if $W_0=0$, the superpotential
has an $R$-symmetry under which $\phi$ has a charge $+2/n$.  Although
such an $R$-symmetry is explicitly broken by the constant term $W_0$,
we will be interested in the case that the effect of the breaking is
relatively small.  Then, in the vacuum where $\langle\phi\rangle\neq
0$, quasi Nambu-Goldstone boson, which we call $R$-axion, shows up.

Including the SUSY breaking effect, the zero temperature potential of
the flaton $\phi$ takes the following form :
\begin{equation}
  V = V_0 - m^2|\phi |^2 + 
  (n-3)\left( \frac{A\phi^{n}}{nM^{n-3}} + {\rm h.c.}\right) 
  + \frac{|\phi|^{2(n-1)}}{M^{2(n-3)}}.
\end{equation}
If there are no sources for the $A$-parameter other than $W_0$, we
have $A=m_{3/2}$.  The SUSY breaking mass parameter for $\phi$ is
represented by $-m^2$ and is assumed to be negative.  Hereafter we
take $A$ real and positive.  In addition, we take $A \ll m$ for
simplicity.  The vacuum expectation value (VEV) of $\phi$ is given by
\begin{equation}
  \langle \phi \rangle \equiv v = 
  \left( \frac{mM^{n-3}}{\sqrt{n-1}} \right)^{1/(n-2)}.    \label{v}
\end{equation}
The flaton mass around the minimum is given by $m_\phi =
\sqrt{2(n-2)}m$, and $V_0$ is determined as
\begin{equation}
  V_0 = \frac{n-2}{n-1}m^2 v^2,
\end{equation}
so that the cosmological constant vanishes at the vacuum.  In the
gravity-mediated SUSY breaking model, $m$ is expected to be of the
order of TeV.  In the gauge-mediated SUSY breaking
model~\cite{Giudice:1998bp}, on the other hand, the scalar potential
is slightly more complicated~\cite{de Gouvea:1997tn,Banks:2002sd} and
the parameter dependence changes.  Hereafter, we treat $v$ and
  $m_\phi$ as free parameters without specifying the mechanism of SUSY
  breaking.

Suppose that $\phi$ is trapped at the origin due to the large Hubble
mass term during/after inflation.  In such a period, $Q$ and $\bar{Q}$
are massless and hence they participate to thermal bath.  Thus the
flaton obtains a thermal mass of $m_T^2 \sim k^2 T^2$.  As the cosmic
temperature decreases, the energy density of the Universe is dominated
by the potential energy of $\phi$ and thermal inflation begins.
Thermal inflation ends at the temperature $T_{\rm end}\sim m$ for
$k\sim \mathcal O(1)$.  Then the flaton oscillates around the minimum,
and finally it decays into radiation.  We parametrize the flaton decay
rate as
\begin{equation}
	\Gamma = \frac{c}{4\pi}\frac{m_\phi^3}{v^2},    \label{gamma}
\end{equation}
with a numerical constant $c$, which depends on the decay mode.
If $\phi$ decays into the Higgsino pair via the superpotential of
  $k' \phi H_u H_d$ (where $H_u$ and $H_d$ are the up- and down-type
  Higgses, respectively) with $k'\sim m_\phi/v$, then $c\sim\mathcal
  O(1)$.  On the contrary, $c\sim\mathcal O(10^{-4})$ for the decay
  into gluons via the loop effect. The reheating temperature after
the flaton decay is estimated as\footnote{ 
    The flaton can also decay into the $R$-axion pair.  If the decay into the
    $R$-axion is dominant, the final reheating completes by the decay
    of $R$-axion. This situation is effectively incorporated in the parametrization (\ref{gamma}) 
    by choosing the value of $c$ appropriately.
    In addition, the flatino, the fermionic component of
    the flaton, has a mass of $\sqrt{n-1}m$, and hence the flaton
    cannot decay into the flatino pair.  }
\begin{equation}
  T_{\rm R}=\left( \frac{90}{\pi^2 g_*} \right)^{1/4}\sqrt{\Gamma M_P}
  \simeq 6.2\times 10^{2}{\rm GeV}
  \sqrt{c}
  \left[ \frac{228.75}{g_*(T_{\rm R})} \right]^{1/4}
  \left( \frac{m_\phi}{1{\rm TeV}} \right)^{3/2}
  \left( \frac{10^{10}{\rm GeV}}{v} \right),
  \label{TR}
\end{equation}
where $g_*(T)$ is the effective number of massless degrees of freedom
at the temperature $T$.  Notice that $T_{\rm R}$ is required to be
higher than a few MeV for successful BBN.

Let us estimate the modulus abundance in the presence of thermal
inflation.  We denote the modulus field by $\chi$ and its mass and
energy density by $m_\chi$ and $\rho_\chi$.  Thermal inflation
efficiently dilutes the moduli through the late-time entropy
production.  The dilution factor is defined as the ratio of the
entropy with and without the additional entropy production at the
flaton decay. It is given by~\cite{Lyth:1995ka,Asaka:1999xd}.
\begin{equation}
\begin{split}
  \Delta = \frac{30}{\pi^2 g_*(T_{\rm end})} 
  \frac{V_0}{T_{\rm end}^3 T_{\rm R}} 
  \simeq 1.3 \times 10^{15} 
  \left[ \frac{228.75}{g_*(T_{\rm end})} \right]
  \left( \frac{1{\rm TeV}}{m} \right)
  \left( \frac{1{\rm GeV}}{T_{\rm R}} \right)
  \left( \frac{v}{10^{10}{\rm GeV}} \right)^2.
\end{split}
\end{equation}
The modulus abundance after the entropy production is
\begin{equation}
  \begin{split}
    \left (\frac{\rho_\chi}{s}\right)^{\rm (prim)} 
    &= \frac{1}{\Delta}\times 
    \frac{1}{8}T_{\rm R}^{(\rm inf)}\left( \frac{\chi_0}{M_P} \right)^2 \\
    & \simeq 1\times 10^{-7}{\rm GeV}\sqrt{c}
    \left( \frac{m}{1{\rm TeV}} \right)^{5/2}
    \left( \frac{10^{10}{\rm GeV}}{v} \right)^3
    \left( \frac{T_{\rm R}^{\rm (inf)}}{10^6{\rm GeV}} \right)
    \left( \frac{\chi_0}{M_P} \right)^2,   \label{mod-prim}
  \end{split}
\end{equation}
where $\chi_0$ is the initial amplitude of the modulus.  Here we have
assumed that the modulus begins to oscillate before the inflaton
decays and also that the inflaton decays before thermal inflation
starts.  Otherwise, the expression becomes more
complicated~\cite{Asaka:1999xd}.  We call this ``primary'' moduli for
the reason discussed below.

There is another contribution to the modulus
oscillation~\cite{Lyth:1995ka}.  Since the potential minimum of the
moduli during thermal inflation may be displaced from the true minimum
due to the Hubble mass correction, the secondary oscillation is
induced after thermal inflation ends.  The amplitude is estimated as
$\delta \chi \sim V_0 \chi_* / (m_\chi^2 M_P^2)$.  Hence the moduli
abundance is estimated to be
\begin{equation}
\begin{split}
	\left (\frac{\rho_\chi}{s}\right)^{\rm (sec)} &
        = \frac{m_\chi^2 (\delta\chi)^2/2}{V_0}\frac{3T_{\rm R}}{4} \\
	& \simeq 1\times 10^{-15}{\rm GeV}\sqrt{c}
	\left( \frac{m}{1{\rm TeV}} \right)^{7/2}
	\left( \frac{1{\rm TeV}}{m_\chi} \right)^2
	\left( \frac{v}{10^{10}{\rm GeV}} \right)
	\left( \frac{\chi_*}{M_P} \right)^2.   \label{mod-sec}
\end{split}
\end{equation}
Here $\chi_*$ is the true minimum of the modulus while the origin is chosen so that the
Hubble mass correction is  given by $H^2\chi^2$.
This is called the ``secondary'' moduli.  The final modulus abundance
is the sum of the primary and secondary moduli.  Which one dominates
the moduli density depends on various parameters.  BBN, cosmic
microwave background, and diffuse $X$- and $\gamma$-ray background
give stringent upper bound on the modulus abundance, depending on its
mass.  Complete analyses on the modulus abundance and its
cosmological effects can be found in Ref.~\cite{Asaka:1999xd}.

\section{Domain wall problem and a solution}

In this section we discuss the problematic DW formation after thermal
inflation, and a solution to the DW problem.

After the decay, the flaton randomly falls into one of the
$n$-degenerate minima.  Since the VEV of $\phi$ spontaneously breaks
the $Z_{n}$-symmetry, DWs are formed after thermal inflation.  The DW
tension is estimated as
\begin{equation}
	\sigma \simeq \frac{V_A}{m_a},
\end{equation}
where
\begin{equation}
	V_A =  2(n-3)A\frac{v^{n}}{nM^{n-3}},
\end{equation}
and
\begin{equation}
	m_a^2 = \frac{n(n-3)}{\sqrt{n-1}} Am
\end{equation}
gives the $R$-axion mass.  DWs obey the scaling low in which about one
DW exists per Hubble horizon as far as the viscosity on the DW is
negligible~\cite{Vilenkin}.\footnote{
	Actually interactions of the DW with $Q$ and $\bar Q$ cause friction, but this effect is found to be insignificant
	for the following analysis.
}
Then DWs begin to dominate the Universe
at
\begin{equation}
	H_{\rm dom} \simeq \frac{\sigma}{M_P^2}.
\end{equation}
This is problematic unless $\sigma \lesssim (1{\rm MeV})^3$, since
otherwise DWs dominate the Universe before the present
epoch~\cite{Zeldovich:1974uw}.  Thus there should be an explicit
$Z_{n}$ breaking term which makes DWs unstable.

Let us denote the $Z_n$-breaking scalar potential by $V_\epsilon$,
whose possible origins will be discussed later.  It generates a bias
for the $n$ minima of the original potential and the degeneracy among
those minima are lifted completely.  DWs are not absolutely stable
under the existence of the bias, and eventually collapse when the bias
energy density becomes comparable to the DW energy
density~\cite{Vilenkin:1981zs}. This happens when
\begin{equation}
	H_{\rm dec} \simeq \frac{V_\epsilon}{\sigma}.
\end{equation}
Therefore, in order for the DWs to collapse before they come to dominate ($H_{\rm dec} \gg H_{\rm dom} $), 
we need the following condition,
\begin{equation}
	V_\epsilon \gg \frac{\sigma^2}{M_P^2}.   \label{Veps_sigma}
\end{equation}
The DW problem is solved if this condition is satisfied.  Now we
describe possible origins of the bias term.

\subsection{Small explicit $Z_n$ breaking term}

Let us introduce the following superpotential
\begin{equation}
  \delta W = \epsilon \frac{\phi^{\ell}}{\ell M^{\ell-3}},   \label{delW}
\end{equation}
in addition to (\ref{W}), where $n$ and $\ell$ are relatively prime
numbers and $\epsilon$ is a small coefficient.  In order for this term
not to change the flaton potential significantly, we require
\begin{equation}
  \epsilon \frac{v^{\ell}}{\ell M^{\ell-3}} \ll  \frac{v^{n}}{n M^{n-3}}
  ~~~\leftrightarrow~~~ \epsilon \ll \frac{\ell}{n} 
  \left( \frac{v}{M} \right)^{n-\ell}.
\end{equation}
Then the following additional scalar potentials are generated,
\begin{equation}
  V_\epsilon = \epsilon A (\ell-3) \frac{\phi^{\ell}}{\ell M^{\ell-3}} + 
  \epsilon \frac{\phi^{n-1}\phi^{*\ell-1}}{M^{n+\ell-6}}
  +{\rm h.c.} .
\end{equation}
The second term dominates for $m>A$ and this yields the bias for the
$n$ minima of the original potential as
\begin{equation}
	V_\epsilon \simeq 2\epsilon \frac{v^{n-\ell-2}}{M^{n+\ell-6}}.
\end{equation}
The original $Z_n$ symmetry is explicitly violated by the bias, and
hence the degeneracy of the $n$ minima of the scalar potential is
broken.  The condition (\ref{Veps_sigma}) is rewritten as
\begin{equation}
  \epsilon \gg \frac{A}{m}\frac{v^2}{M_P^2}
  \left( \frac{v}{M} \right)^{n-\ell}.
\end{equation}
If this condition is satisfied, DWs collapse before they come to
dominate the Universe and there is no DW problem.

\subsection{QCD instanton effect}

Rather simple scenario to break the $Z_n$ symmetry is to use the
  QCD instanton effects~\cite{Preskill:1991kd}.  This is economical in
  the sense that the term $\phi Q\bar Q$ in (\ref{W}), which is needed
  for giving rise to a thermal mass for the flaton field, also works as a source of the bias for
  eliminating the DWs.\footnote{ This solution may be incompatible
    with the Peccei-Quinn mechanism~\cite{Peccei:1977hh} for solving
    the strong CP problem~\cite{Preskill:1991kd,Abel:1995wk}.  In that
    case, we may arrange the model so that the heavy quarks $Q (\bar
    Q)$ have a charge of {\it hidden} QCD and its scale is around
    GeV-TeV.  }

The $Z_n$ symmetry in (\ref{W}) has an anomaly for SU(3)$_{\rm QCD}$.
Consequently, the QCD instanton effect works as a bias, which lifts
the classical degeneracy of the $n$ vacua.  The bias due to this
effect is estimated to be~\cite{Kim:1986ax}
\begin{equation}
  V_\epsilon \simeq f_\pi^2 m_\pi^2 \frac{m_u m_d}{(m_u + m_d)^2},
\end{equation}
for temperature below $\Lambda_{\rm QCD}\sim 200$MeV, where $f_\pi$
and $m_\pi$ are the pion decay constant and mass, and $m_u (m_d)$ is
the current mass of the up (down) quark.  At higher temperature, this
effect is suppressed by high powers of $T$~\cite{Gross:1980br}, and
hence the bias due to the QCD instanton effect is only turned on at
$T\sim \Lambda_{\rm QCD}$~\cite{Preskill:1991kd, Abel:1995wk,
  Ibe:2004gh, Riva:2010jm}.

We need two conditions for this mechanism to work successfully.  One
is Eq.~(\ref{Veps_sigma}) for DWs collapse due to the bias. The
other is $H_{\rm dec} \lesssim \Lambda_{\rm QCD}^4/(T_{\rm R}^2M_P)$,
since otherwise the QCD instanton effect is highly suppressed and it
is not suitable for the bias.\footnote{ The latter
  condition is based on the assumption that the Universe is flaton
  dominated at the QCD phase transition.  If the Universe is radiation
  dominant at the QCD phase transition, the condition is severer.  }
These two conditions are written as
\begin{equation}
	T_{\rm R}^2 M_P < \sigma < \Lambda_{\rm QCD}^2 M_P
\end{equation}
for $T_{\rm R}<\Lambda_{\rm QCD}$.  Therefore, for example, $M\sim
10^{15}$GeV and $A=m_{3/2} \sim m\sim 1$GeV are good choices for
$n=4$. In this choice we have $H_{\rm dec}/H_{\rm dom} \sim \mathcal
O(1)$.  As will be seen, this predicts observable GWs.

It should be noticed that $M$ cannot take an arbitrary value in order
to solve the cosmological moduli problem.  As seen in
Eqs.~(\ref{mod-prim}) and (\ref{mod-sec}), the moduli abundance
crucially depends on $v$, which is determined by the cutoff scale $M$.
Comparing with the BBN bound $\rho_\chi/s \lesssim 10^{-14}$GeV for
the moduli of $10{\rm GeV}\lesssim m_\chi \lesssim
1$TeV~\cite{Kawasaki:2004qu}, $M \gtrsim 10^{15}$GeV for $n=4$ may be
allowed for moderate parameter choices.

\section{Gravitational waves from collapsing domain walls}

Now we are at the position to discuss the main subject of this paper,
which is the GW production from DWs.  We will show that the GWs from
the DWs after thermal inflation may be observable by future experiments.

The DW network has complicated structure.  Large amount of energy is
localized around DWs, which becomes the source GWs.  As a result,
considerable amount of GWs can be emitted by the relativistic motion
of DWs.  Frequencies of the GWs correspond to typical scales of the DW
motion, which ranges from the Hubble scale to the size of DW width.

Detailed calculations of the GW spectrum from DWs using three
dimensional lattice simulation were performed in
Refs.~\cite{Gleiser:1998na,Hiramatsu:2010yz,Kawasaki:2011vv}.
According to recent studies~\cite{Hiramatsu:2010yz,Kawasaki:2011vv} ,
GWs have rather broad spectrum which extends from the horizon size at
the DW collapse for lower frequency side to the DW width for the
higher frequency side.  The GW spectrum, in terms of $\Omega_{\rm
  GW}(f)\equiv (d\rho_{\rm GW}/d\ln f)/\rho_{c0}$ (where $\rho_{\rm
  GW}(f)$ is the GW energy density with frequency $f$ measured at
present and $\rho_{c0}$ is the present critical energy density) is
almost flat between these frequencies.  The energy density of
gravitational waves at the collapse of DWs can be estimated as
\cite{Takahashi:2008mu}
\begin{equation}
  \rho_{\rm GW}(H_{\rm dec}) \simeq 
  G_N \frac{M_{\rm DW}^2}{H_{\rm dec}^{-1}} \frac{1}{H_{\rm dec}^{-3}} 
  \sim \frac{\sigma^2}{M_P^2},
\end{equation}
where $G_N$ is the Newton constant.  Here we have substituted $M_{\rm
  DW} = \sigma H_{\rm dec}^{-2}$.  Then, $\Omega_{\rm GW}$ is given by
\begin{eqnarray}
  \displaystyle 
  \Omega_{\rm GW}(f) \simeq \Omega_{\rm r} 
  \left[ \frac{g_*(T_{\rm dec})}{g_{*0}} \right]
  \left[ \frac{g_{*s0}}{g_{*s}(T_{\rm dec})} \right]^{4/3}
  \left( \frac{H_{\rm dom}}{H_{\rm dec}} \right)^2 
  & {\rm ~~for~~} H_{\rm R} > H_{\rm dec},   \\
  \displaystyle
  \Omega_{\rm GW}(f) \simeq \Omega_{\rm r} 
  \left[ \frac{g_*(T_{\rm dec})}{g_{*0}} \right]
  \left[ \frac{g_{*s0}}{g_{*s}(T_{\rm dec})} \right]^{4/3}
  \left( \frac{H_{\rm dom}}{H_{\rm dec}} \right)^2 
  \left( \frac{H_{\rm R}}{H_{\rm dec}} \right)^{2/3}
  & {\rm ~~for~~} H_{\rm R} < H_{\rm dec},
\end{eqnarray}
for $f_{\rm edge}<f<f_{\rm peak}$ defined below, where $\Omega_{\rm r}
= 8.5\times 10^{-5}$ is the present radiation energy with three
massless neutrino species density. Numerically we obtain
\begin{eqnarray}
  \displaystyle 
  \Omega_{\rm GW}(f) \simeq 2\times 10^{-5}
  \left[ \frac{228.75}{g_*(T_{\rm dec})} \right]^{1/3} 
  \left( \frac{H_{\rm dom}}{H_{\rm dec}} \right)^2 
  & {\rm ~~for~~} H_{\rm R} > H_{\rm dec},   \\
  \displaystyle
  \Omega_{\rm GW}(f) \simeq 2\times 10^{-5}
  \left[ \frac{228.75}{g_*(T_{\rm dec})} \right]^{1/3}
  \left( \frac{H_{\rm dom}}{H_{\rm dec}} \right)^2 
  \left( \frac{H_{\rm R}}{H_{\rm dec}} \right)^{2/3}
  & {\rm ~~for~~} H_{\rm R} < H_{\rm dec},
\end{eqnarray}
for $f_{\rm edge}<f<f_{\rm peak}$.  The typical frequency at the lower
side, corresponding to the horizon size at the DW collapse, called the
``edge'' frequency ($f_{\rm edge}$) in Ref.~\cite{Hiramatsu:2010yz},
is given by
\begin{equation}
  f_{\rm edge} = \frac{H_{\rm dec}}{2\pi}\frac{a(H_{\rm dec})}{a_0},
\end{equation}
which is estimated as
\begin{eqnarray}
  \displaystyle 
  f_{\rm edge} \simeq 3\times 10^{-5}{\rm Hz} 
  \left[ \frac{g_*(T_{\rm dec})}{228.75} \right]^{1/6}  
  \left( \frac{T_{\rm dec}}{1{\rm TeV}} \right) 
  & {\rm ~~for~~} H_{\rm R} > H_{\rm dec},   \\
  \displaystyle
  f_{\rm edge} \simeq 3\times 10^{-5}{\rm Hz}  
  \left[ \frac{g_*(T_{\rm R})}{228.75} \right]^{1/6} 
  \left( \frac{T_{\rm R}}{1{\rm TeV}} \right) 
  \left( \frac{H_{\rm dec}}{H_{\rm R}} \right)^{1/3}
  & {\rm ~~for~~} H_{\rm R} < H_{\rm dec},
\end{eqnarray}
where $T_{\rm dec}$ is the temperature at which the DWs collapse.  On
the other hand, the ``peak'' frequency ($f_{\rm
  peak}$)~\cite{Hiramatsu:2010yz}, corresponding to the DW width at
the collapse redshifted to the present time, is given by
\begin{equation}
  f_{\rm peak} = \frac{w^{-1}}{2\pi}\frac{a(H_{\rm dec})}{a_0},
\end{equation}
where $w\simeq m_a^{-1}$ is the DW width. This is estimated as
\begin{eqnarray}
  \displaystyle 
  f_{\rm peak} \simeq 1\times 10^{10}{\rm Hz} 
  \left[ \frac{228.75}{g_*(T_{\rm dec})} \right]^{1/3}
  \left( \frac{m_a}{1{\rm TeV}} \right)
  \left( \frac{1{\rm TeV}}{T_{\rm dec}} \right)
  & {\rm ~~for~~} H_{\rm R} > H_{\rm dec},   \\
  \displaystyle
  f_{\rm peak} \simeq 1\times 10^{10}{\rm Hz}  
  \left[ \frac{228.75}{g_*(T_{\rm R})} \right]^{1/3} 
  \left( \frac{m_a}{1{\rm TeV}} \right)
  \left( \frac{1{\rm TeV}}{T_{\rm R}} \right) 
  \left( \frac{H_{\rm R}}{H_{\rm dec}} \right)^{2/3}
  & {\rm ~~for~~} H_{\rm R} < H_{\rm dec}.
\end{eqnarray}
We approximate that the GW spectrum is flat between $f_{\rm
    edge}$ and $f_{\rm peak}$, although the simulation shows a bit
preference for the blue spectrum~\cite{Hiramatsu:2010yz}.  In that
sense, our estimate of the GW is conservative.  It may be within the
observable range of GW detectors such as pulsar timing array or space
laser interferometers~\cite{Maggiore:1999vm}.

\begin{figure}[htbp]
\begin{center}
\includegraphics[scale=1.2]{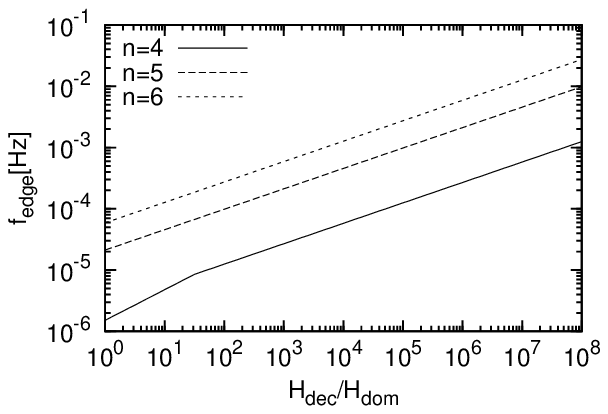}
\includegraphics[scale=1.2]{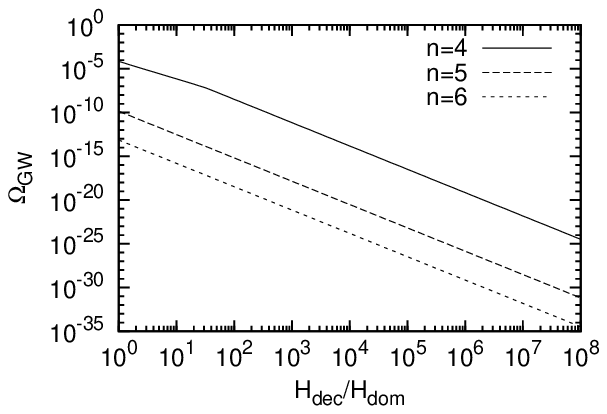}
\caption{
  The typical GW frequency $f_{\rm edge}$ (left) and $\Omega_{\rm GW}$ (right)
  as a function of $H_{\rm dec}/H_{\rm dom}$.
  We have taken $m=1$TeV, $A=100$GeV, $c=1$, $M=M_P$ and $n=4,5,6$.
}
\label{fig:OGW}
\end{center}
\end{figure}

Fig.~\ref{fig:OGW} shows the edge frequency $f_{\rm edge}$ (left) and
$\Omega_{\rm GW}$ (right) as a function of $H_{\rm dec}/H_{\rm dom}$,
which parameterizes the epoch of the collapse of DWs.  
Here we have taken $m=1$TeV, $A=100$GeV,
$c=1$, $M=M_P$ and $n=4,5,6$.  
It is clear that the GW energy density becomes larger for DWs
that collapse closer to the epoch of DW domination.

\begin{figure}[htbp]
\begin{center}
\includegraphics[scale=0.8]{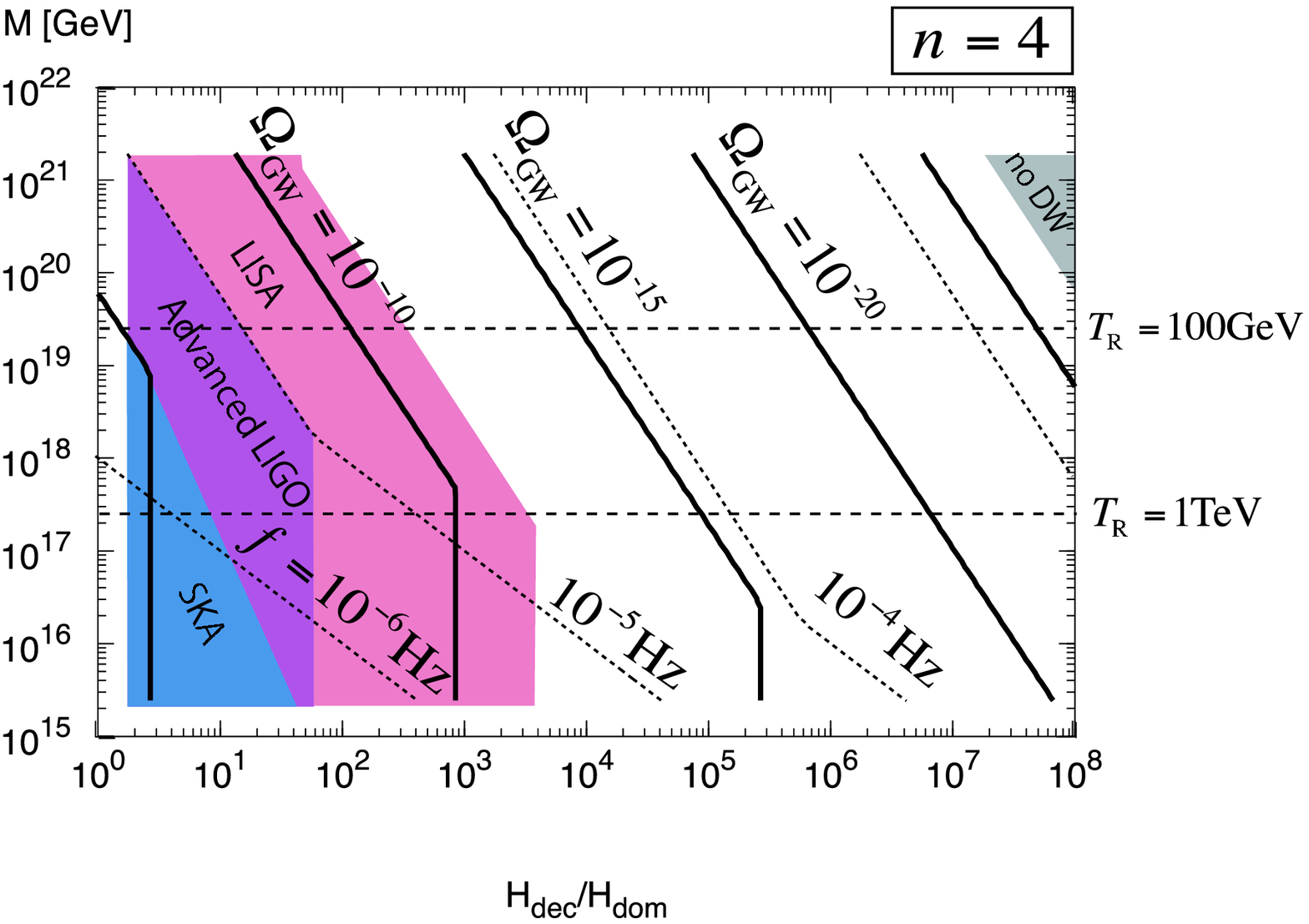}
\includegraphics[scale=0.8]{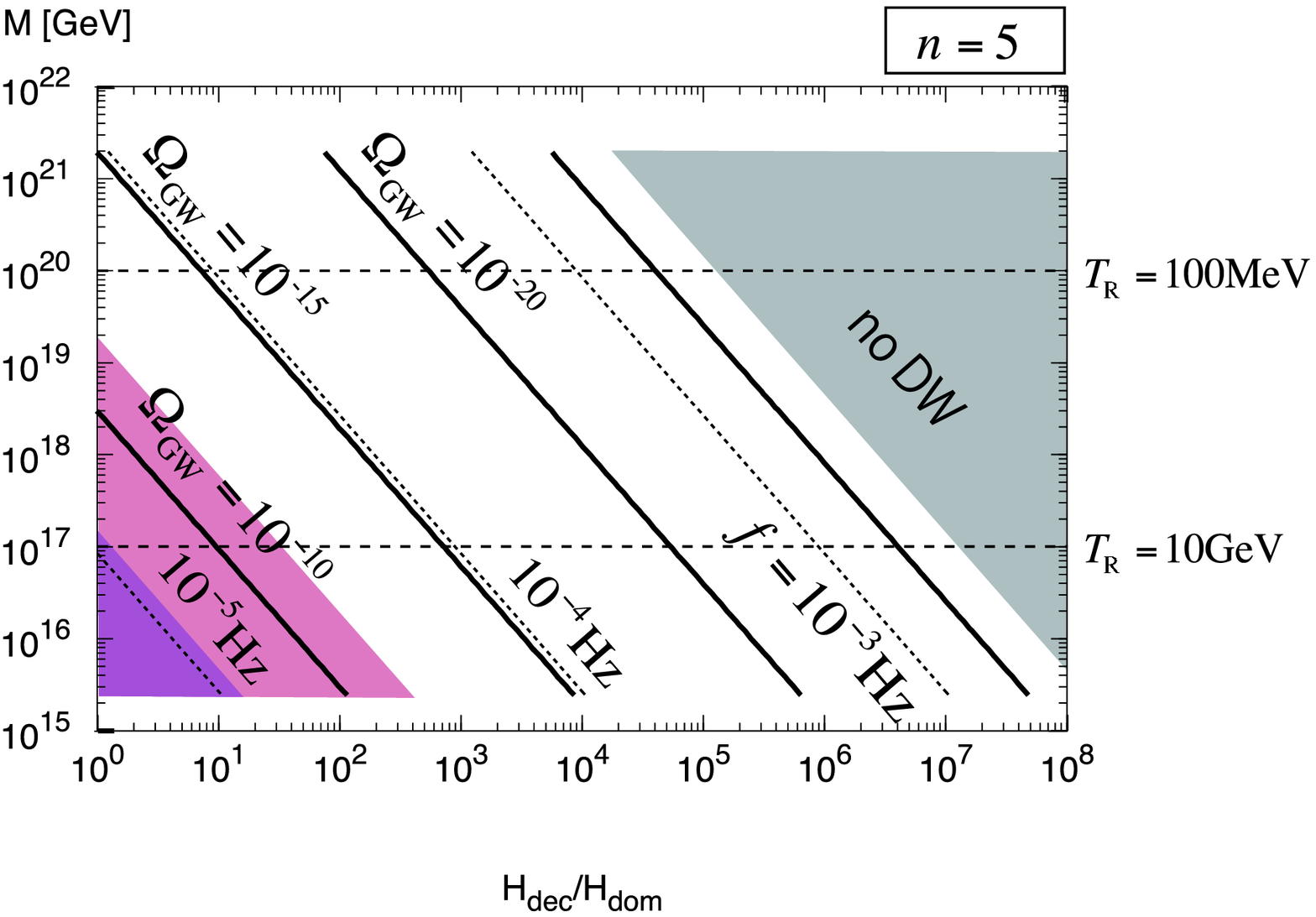}
\caption{
  Contours of $\Omega_{\rm GW}$ (solid line), $f_{\rm edge}$ (dotted line) and $T_{\rm R}$ (dashed line)
  for $n=4$ (top) and $n=5$ (bottom).
  The regions explored by LISA, SKA and advanced LIGO are also shown.
  DECIGO may cover the whole parameter region with 
  $\Omega_{\rm GW} \gtrsim 10^{-18}$.
  Here we have taken $m=1$TeV, $A=100$GeV and $c=1$.
 }
\label{fig:contour}
\end{center}
\end{figure}

Fig.~\ref{fig:contour} shows contours of $\Omega_{\rm GW}$, $f_{\rm
  edge}$ and $T_{\rm R}$ for $n=4$ (top) and $n=5$ (bottom).  
  In the gray region denoted by ``no DW'', the bias is so large, i.e. $H_{\rm
  dec} > \sqrt{V_0}/M_P$, that no DW formation is expected.  The
current limits from pulsar timing
experiments~\cite{Kaspi:1994hp,Lommen:2002je,vanHaasteren:2011ni} and
the LIGO experiment~\cite{Abbott:2009ws} do not give stringent
constraints on the model.  In the same figure, we also show the
  prospects of the discovery reaches of up-comming projects.
Projects such as SKA~\cite{Kramer:2004rwa} will probe some parameter
regions.  Ground based GW detectors such as the advanced
LIGO~\cite{AdvLIGO} and LCGT\cite{Kuroda:2010zz} are also sensitive to
the higher frequency GWs of $f \gtrsim 100$~Hz.  Here a correlation
analysis of 1 year is assumed.  Space based GW detector such as LISA~\cite{Cornish:2001bb}
will also be sensitive to the GWs from DW collapse for wider parameter
regions.  Sensitivities are found in Ref.~\cite{Sathyaprakash:2009xs}.
DECIGO~\cite{Seto:2001qf} may cover the whole parameter
region with $\Omega_{\rm GW} \gtrsim 10^{-18}$.  The scenario with the
QCD instanton as a bias term corresponds to the lower-left edge of
Fig.~\ref{fig:contour} and hence future pulsar timing and/or
space-borne GW experiments will be able to detect GWs.

\section{Discussion and Conclusions }

Some comments are in order.  It was proposed that the first order
phase transition triggers the end of thermal inflation, and
correspondingly bubbles of true vacuum are
created~\cite{Easther:2008sx}.  Bubbles expand into the surrounding
false vacuum regions and collide with each other, generating GWs with
observable level~\cite{Easther:2008sx}.  This predicts higher
frequency GWs than studied in this paper and hence will provide
further information on the thermal inflation model.

We also comment on possible origin of the baryon asymmetry in the
model.  Since thermal inflation dilutes away the pre-existing baryon
asymmetry, as well as the moduli, we need some mechanisms to create
baryon asymmetry after thermal inflation.  In this regard, a variant
type of the Affleck-Dine mechanism~\cite{Affleck:1984fy} works after
thermal inflation~\cite{Stewart:1996ai, Jeong:2004hy, Kawasaki:2006py,
  Felder:2007iz, Kim:2008yu}.

One might notice that the heavy quraks $Q$ and $\bar{Q}$ may be stable
and cosmologically relevant~\cite{Hui:1998dc}.  Since they are once in
thermal equilibrium during thermal inflation, its
abundance just after the flaton decay is given by
\begin{equation}
\begin{split}
	Y_Q &\equiv \frac{n_Q}{s}\sim \frac{T_{\rm end}^3 T_{\rm R}}{m^2 v^2} 
	\sim 10^{-17}\left( \frac{m_\phi}{1{\rm TeV}} \right)
	\left( \frac{T_{\rm R}}{1{\rm GeV}} \right)
	\left( \frac{10^{10}{\rm GeV}}{v} \right)^2.
\end{split}
\end{equation}
Even if it is less than the DM abundance, BBN constraints on the
strongly-interacting relic particles are much more
stringent~\cite{Kusakabe:2009jt,Kawasaki:2010yh}.\footnote{ The
  enhancement of the annihilation cross section due to the $R$-hadron
  formation~\cite{Kang:2006yd} may help the situation if the flaton
  decays after the QCD phase transition.  } This constraint can easily
be evaded if $Q$ and $\bar{Q}$ are in complete multiplets of the GUT
group, which is natural from the viewpoint of gauge coupling
unification.  If $Q$ and $\bar{Q}$ are embedded into fundamental and
anti-fundamental representation of SU(5)$_{\rm GUT}$, for example,
they should be acompanied by non-colored components (denoted by
$L$ and $\bar{L}$).  The renormalization group effect makes the
colored components heavier than non-colored ones at the low energy.
Then, $Q$ and $\bar{Q}$ decay into $L$ and $\bar{L}$ via dimension
five operators suppressed by the GUT scale.  The decay rate is
estimated to be $\Gamma \sim m_Q^3 / M_{\rm GUT}^2$, where $M_{\rm
  GUT}\sim 10^{16}{\rm GeV}$ is the GUT scale.  The decay rate is
large enough so that colored components decay before BBN.  Similarly,
$L$ splits into the electrically charged and neutral components, where
the former is heavier and decays into the latter.  Notice that
  the neutral component may be a candidate of superheavy
  DM.\footnote{The constraints from direct detection experiments
      can be avoided if the mass of the neutral component is heavier
      than $\sim 10^{5}\ {\rm GeV}$~\cite{Srednicki:1986vj}.} 

Let us consider how robust the DW formation after thermal inflation
is.  In the simplest model, the flatness of the flaton potential is
ensured by a $Z_n$ symmetry as in (\ref{W}).  In this case DWs
necessarily appear.  On the other hand, one can impose a global U(1)
symmetry to make the flaton potential flat.  The flaton potential is
stabilized by introducing another singlet scalar with some U(1) charge
or by taking account of the radiative correction to the flaton soft
mass~\cite{Kim:2008yu, Asaka:1998ns, Chun:2000jr, Choi:2009qd,
  Park:2010qd, Choi:2011rs}.  Then, the spontaneous breaking of the
U(1) symmetry due to the VEV of the flaton results in the formation of
cosmic strings instead of DWs.  Even so, the global U(1) symmetry may
be anomalous and may not be a good symmetry at the quantum level.  If
only a discrete subgroup $Z_N$ of the U(1) symmetry remains at the
quantum level, DWs still exist~\cite{Sikivie:1982qv}.  In particular,
if $N\geq 2$, the DW problem still exists.  On the contrary, if $N=1$,
DWs are bounded by strings and they disappear just after the
formation~\cite{Vilenkin:1982ks, Vachaspati:1984dz, Nagasawa:1994qu,
  Chang:1998tb, Hiramatsu:2010yn}.

It is also possible that the U(1), which guarantees the flatness of
the potential, is a gauge symmetry~\cite{Lyth:1995hj}.  The flaton
takes the role of the Higgs field that gives a mass for the U(1) gauge
boson.  Then, the U(1) should be an exact symmetry and DW does not
exist.  Instead, cosmic strings are formed which are less harmful than
DWs.  Cosmic strings formed after thermal inflation has a width of the
TeV scale, while the tension can be much larger.  This kind of thick
strings have characteristic cosmological implications as studied in
Refs.~\cite{Cui:2007js,Kawasaki:2011dp}.

Finally, we comment on another possible source of GWs, which is
  the DW related to the breaking of Peccei-Quinn
  symmetry~\cite{Peccei:1977hh}.  In order to avoid the DW problem, it
  is often the case that axion models with $N=1$ is considered.  In
  such a case, DWs are surrounded by cosmic strings as we have
  mentioned; then, DWs disappear just after the production.  This
  happens when $T\sim\Lambda_{\rm QCD}$, so we expect that the GWs
  with the frequency of $f\sim 10^{-8}\ {\rm Hz}$ are emitted from the
  dynamics of axionic DWs.  We have estimated the amount of such GWs,
  and found that $\Omega_{\rm GW}$ is a few order of magnitude smaller
  than the sensitivity of the future pulsar timing experiments.

To summarize, a class of thermal inflation models is inevitably
associated with DW formation, which reintroduces a cosmological
disaster.  Thus we need a mechanism to make DWs unstable.  If the DW
energy density at the epoch of its collapse is large enough, GWs
produced by the DW collapse may be observable in future experiments.
We have pointed out that such a mechanism is naturally built-in
  in the simplest model where the flaton couples to extra quarks as in
  (\ref{W}); in such a model, the $Z_n$ symmetry has an anomaly for
  SU(3)$_{\rm QCD}$ and the QCD instanton effects serve as a bias for
  lifting the degeneracy among $n$ minima.  In this case the GW
  amplitude becomes so large that it can be well within the reach of
  future experiments. 

\begin{acknowledgments}

  K.N. would like to thank M.~Kawasaki and F.~Takahashi for useful conversations.
  This work is supported by Grant-in-Aid for Scientific research from
  the Ministry of Education, Science, Sports, and Culture (MEXT),
  Japan, No.\ 22540263 (T.M.), No.\ 22244021 (T.M.), No.\ 21111006
  (K.N.), and No.\ 22244030 (K.N.).

\end{acknowledgments}

  
 
 \end{document}